\documentclass[12pt]{article}
\usepackage{epsfig}
\def\be{\begin{equation}}
\def\ee{\end{equation}}
\def\ba{\begin{eqnarray}}
\def\ea{\end{eqnarray}}
\def\ga{\mathrel{\raise.3ex\hbox{$>$\kern-.75em\lower1ex\hbox{$\sim$}}}}
\def\la{\mathrel{\raise.3ex\hbox{$<$\kern-.75em\lower1ex\hbox{$\sim$}}}}

\begin{document}
\baselineskip=16pt
\begin{titlepage}
\begin{center}

\vspace{0.5cm}
{\Large \bf
Non-minimally Coupled Tachyon and Inflation
}\\
\vspace{10mm}

Yun-Song Piao$^{a,c}$, Qing-Guo Huang$^{c}$, Xinmin Zhang$^{a}$ and
Yuan-Zhong Zhang$^{b,c}$ \\
\vspace{6mm}
{\footnotesize{\it
 $^a$Institute of High Energy Physics, Chinese
     Academy of Sciences, P.O. Box 918(4), Beijing 100039, China\\
 $^b$CCAST (World Lab.), P.O. Box 8730, Beijing 100080\\
 $^c$Institute of Theoretical Physics, Chinese Academy of Sciences,
      P.O. Box 2735, Beijing 100080, China \footnote{Email address:
      yspiao@itp.ac.cn, huangqg@itp.ac.cn}\\}}

\vspace*{5mm}
\normalsize
\smallskip
\medskip
\smallskip
\end{center}
\vskip0.6in
\centerline{\large\bf Abstract}
 {
In this paper, we consider a model of tachyon with a non-minimal 
coupling to gravity and study its cosmological effects. 
Regarding inflation
we will show that for a specific coupling of tachyon to gravity
this 
model satisfies
observations and solves various problems
which exist
in the single and multi tachyon inflation models.
}
\vspace*{2mm}
\end{titlepage}

Recently there has been a lot of studies on the 
tachyon cosmology  
\cite{S, GMP, FC, CGJP, KL, PC}. 
Regarding inflation the unconventional form of the tachyonic
action brings some new features and makes it different from that with a 
normal scalar field, however as pointed out firstly by
Kofman and Linde \cite{KL} that the tachyonic inflations suffer from serious 
difficulties. Authors of Ref. \cite{PC} considered a model of multi 
tachyonic 
inflation, which solves some of the problems in the single tachyon model. 
But still for the string scale $M_s
\leq H$ as required in these
inflation models implies that the size of the de-Sitter horizon
be smaller than the string length {\it i.e.} ${1\over H} \leq
l_s$, which makes it invalid to describe the tachyon condensation
by using an effective field theory \cite{KL}.

In all of these studies, the tachyon action is usually the DBI action with 
the
minimal coupling between the tachyon and gravity \cite{S, G, Se}.
In this paper, we propose a model of tachyon which couples to gravity 
non-minimally and study its cosmology. 
We will show that for a specific form of non-minimal coupling of the 
tachyon to gravity this model satisfies observations and overcomes all of 
the problems which exist in the single and multi tachyon inflation models.
Studies
in the bosonic string  
or for non-BPS brane indicate the possibility on 
the non-minimal couplings between tachyon and gravity 
\footnote{We thank A.A. Tseytlin for communication
about this with us.}\cite{ST}.

The model under consideration is described by an 4D 
effective action of tachyon non-minimally coupled to gravity
 as follows
\be
S= \frac{1}{2\kappa^2} \int d^4 x \sqrt{-g} f(T) R
  +\int d^4 x \sqrt{-g} V(T)
   \sqrt{1+\alpha^\prime g^{\mu\nu} \partial_\mu T \partial_\nu T , }
\ee
where $\kappa^2 ={1\over M_p^2}$ sets the 4D Planck scale and
the tachyon potential around $T=0$ is given by $V(T)=\tau_3 e^{-T^2}$.
$f(T)$ is a function of tachyon field $T$, which corresponds to the 
minimal coupling of tachyon to gravity when $f(T) = 1$.
The tension of the non-BPS brane is
\be
\tau_3 ={\sqrt{2} M_s^4\over (2\pi)^3 g_s} ,
\label{tau3}
\ee
with
$g_s$ being the string coupling and $M_s = l^{-1}_s ={1\over
\sqrt{\alpha^\prime}}$
the fundamental string mass and length scales. The Planck
mass in this model is given by the dimensional reduction
\be
M_p^2 ={v M_s^2\over g_s^2} ,
\label{mp}
\ee
where $v=(M_s r)^6$, $r$ is the radius of the compactification.
The 4D effective theory is applicable only if $v \gg 1$.

Following \cite{M}, we perform a conformal transformation
$g_{\mu\nu}(x) \rightarrow f(T)g_{\mu\nu}(x)$, and obtain that
\be
S= \frac{1}{2\kappa^2} \int d^4 x \sqrt{-g}
\left(R+ {3\over 2}({f^\prime\over f})^2 g^{\mu\nu}\partial_{\mu}
T\partial_{\nu}T\right)
  +\int d^4 x \sqrt{-g} \tilde{V} (T)
   \sqrt{1+\alpha^\prime f(T) g^{\mu\nu} \partial_\mu
   T \partial_\nu T ,}
\label{s}
\ee
 in
Einstein frame, where $\tilde{V}(T)=\frac{V(T)}{f^2(T)}$ is the
tachyon effective potential in this frame. For a 
spatially homogeneous but
time-dependent tachyon field, the Hubble equation is given by \be
H^2=\frac{\kappa^2}{3}
    \frac{\tilde{V}(T)}{\sqrt{1-\alpha^\prime f(T) {\dot T}^2}}
    +{1\over 4}\left({f^\prime\over f}\right)^2 {\dot T}^2 ,
\label{h}
\ee
and the equation of motion for tachyon is
\ba
&&\left(\frac{1}{1-\alpha^\prime f(T) {\dot T}^2}+{3\over 2}
({f^\prime\over f})^2 {M_p^2\over \alpha^\prime f {\tilde V}}
\sqrt{1-\alpha^\prime f {\dot T}^2}\right) \ddot T
+\left(1+{3\over 2}({f^\prime\over f})^2 {M_p^2\over
\alpha^\prime f {\tilde V}}\sqrt{1-\alpha^\prime f
{\dot T}^2}\right)3 H \dot T \nonumber\\
&&+\frac{1}{\alpha^\prime}
 \frac{{\tilde V}^\prime}{\tilde{V} f}
+\frac{1}{2}
\left( \frac{1}{1-\alpha^\prime f(T) {\dot T}^2}+{3 M_p^2\over
\alpha^\prime f {\tilde V}}{f^{\prime\prime} f-f{^\prime 2}
\over f^2}\sqrt{1-\alpha^\prime f {\dot T}^2}\right)
 \left(\frac{f^\prime}{f}\right){\dot T}^2
=0 .
\label{t1}
\ea
When $f(T)=1$, we have checked that the (\ref{h}) and (\ref{t1})
agree with those for a minimally coupled tachyon.  
For $f(T)\neq 1$, however the potential is rescaled to be ${\tilde V}(T)$,
besides 
there are more terms 
proportional to ${\dot T}^2$ and
$\left({f^\prime\over f}\right)^2$ in (\ref{t1}),
which can be neglected consistently when
$\delta \equiv {M_p^2\over \alpha^\prime f {\tilde V}}
\left({f^\prime\over f}\right)^2 \ll 1$.

In the following, we study the inflationary solution of
this model.
The slow-rolling parameters are given by
\be
\epsilon=\frac{M_p^2}{\alpha^\prime }
         \frac{{\tilde V}^{\prime 2}}{{\tilde V}^3 f } ,
\ee
\be
\eta=\frac{M_p^2}{\alpha^\prime }
     \frac{{\tilde V}^{\prime\prime}}{{\tilde V}^2 f } .
     \label{eta}
\ee
In the slow-rolling approximation $\epsilon, \eta\ll 1$ and
$\delta\ll 1$, the Hubble equation and the
equation of motion for the rolling tachyon can be expressed as
\be
H^2=\frac{{\kappa^2 \tilde V}}{3},
\label{h2}
\ee
\be
3H{\dot T}+(\alpha^\prime {\tilde V} f(T))^{-1}{\tilde V}^\prime=0 .
\ee

To solve Eqs.(9) and (10) we need the specific forms of $V(T)$ and $f(T)$.
Assuming
that inflation starts near the top of the tachyon
potential, {\it i.e.} around $T=0$,
we take the tachyon potential 
$V(T)=\tau_3 e^{-T^2}$ which valids around the maximum and has a maximum 
at $T=0$. For
$f(T)$ we expand it about $T=0$ as follows \be
f(T)=1+\sum_i c_i T^{2i} \label{ft}, \ee 
where $c_i$ are coefficients.
 Thus \be \tau_3^{-1}{\tilde
V}(T) = \tau_3^{-1}{V(T)\over f^2 (T)} =1-(1+2c_1)T^2+({1\over
2}+2c_1- 2c_2-c_1^2)T^4 +{\cal O}(T^6). \label{vt} \ee Applying the
slow-rolling condition $\eta <1$, we obtain \be g_s > \frac{(2
\pi)^3}{\sqrt{2}} v \left((1+2c_1)+{\cal O}(1)T^2+\cdots\right).
\label{gs11} \ee 
One can see that when $c_1\neq -{1\over 2}$, the result
is similar to that for a minimally coupled tachyon. 
For $c_1=
-{1\over 2}$, however the leading term proportional to $T^2$ 
disappears, correspondingly the "effective potential" of tachyon $
\tau_3^{-1}{\tilde
V}(T) $ becomes flatter. In this case, 
 Eq.(13) becomes
 \be
g_s > 10^2 v T^2 \label{gs1}. \ee 
We have also checked that 
when $c_1= -{1\over 2}$, $\delta\simeq \eta$ indeed a small parameter.
This justifies 
our
assumption above of $\delta\ll 1$ during inflation, so that we can 
safely drop them off in the calculation.

The amplitude of the gravitational waves produced during inflation
is \be {\cal P}_g\sim\frac{H}{M_p} < 3.6 \times 10^{-5} .\label{gra}
\ee Substituting (\ref{h2}) and (\ref{mp}) into (\ref{gra}), we
have \be g_s^3 < 7 \times 10^{-7} v^2 . \label{gs2} \ee Then
considering (\ref{gs1}) and (\ref{gs2}), we obtain \be v T^6 <
10^{-13} \label{v}. \ee  
One can see that for $v\gg 1$,
when $T <10^{-3}$, the condition (\ref{v}) is satisfied, and in
the meantime, combining (\ref{eta}) and (\ref{h2}), we have
$H^{-1} T^2 <l_s$. As is seen in the following, in this model, the
string length is far less than the cosmological horizon during
inflation, which means that the main problem \cite{KL} troubling
the single and multi tachyon inflation models is overcome.

The number of e-folds during inflation is \be N=\int H dt \simeq
-\int_{T_{60}}^{T_{end}} {\alpha^\prime f {\tilde V}(T)\over
{\tilde V}^\prime (T)} H^2 dT . \label{n} \ee 
In Eq.(18)
 $T_{60}$ is
the field value corresponding to $N\simeq 60$ as required when the
COBE scale exits the Hubble radius, and $T_{end}$ is the field
value at which inflation ends, which is determined by $\eta \sim
1$. In our model, \be T_{60}\sim 0.1 T_{end}\sim {1\over
10^2}\sqrt{{g_s\over v}} . \ee Following the definition of the
amplitudes of the density perturbation in Ref \cite{SS}, we have
\be {\cal P}_s \sim {H^2\over \sqrt{\alpha^\prime {\tilde V} f(T)}
{\dot T}}. \ee Thus \be {\cal P}_s \sim {g_s^2\over 10^2 v^{{3\over
2}} T_{60}^3} \sim 10^3 g_s^{{1\over 2}}. \ee Therefore for ${\cal
P}_s \sim 10^{-5}$, we find that $g_s \sim 10^{-16}$ and $M_s\sim
10^3$Gev, $H^{-1}\sim 10^{26} M_p^{-1}\gg l_s\sim 10^{16}M_p^{-1}$
during inflation. In this case, the string scale is about Tev.

Before concluding we should point out that the inflationary solution of 
Eq.(4) can also be solved by firstly
expanding
$\sqrt{1-\alpha^\prime f(T) {\dot T}^2 }$ in terms of ${\dot T}^2$, then 
renormalizing
the kinetic energy term by a re-definition of the
tachyon field. With this approach we obtain the same results as 
those present above.

In summary, we have considered in this paper a model of tachyon which 
couples to 
gravity non-minimally and studied its cosmological effects.
For a specific coupling of tachyon to gravity 
$f(T) = 1 - \frac{1}{2} T^2$ around $T= 0$  
we have studies the inflationary solution, and found that it
makes the tachyon potential flatter, an successful inflation
is realized. The string length in
our model is far less than the cosmological horizon during
inflation, which makes it feasible to describe the tachyon
condensation by using an effective field theory, and thus the main
problem troubling the single and multi tachyon inflation models is
overcome. Finally, we would like to mention that the usual
reheating mechanism is not feasible \cite{KL} since the tachyon
does not oscillate during the decay of non-BPS branes. There are
some discussions about the tachyon reheating \cite {STW}. But
recently, some studies have pointed out that as the tachyon
evolves into the late-time, the strength of coupling to the closed string
increase \cite{GHY}. These results motivate us
to expect that the tachyon could emit closed string radiation
\cite{GS, D}, such as graviton and dilaton, into the bulk and
eventually settles in the finite minimum.

This work implies that the inflation and
cosmological applications of tachyon with non-minimal couplings
to gravity
may have more fruitful phenomena, which is worth
studying further.

\textbf{Acknowledgments}

We would like to thank Miao Li, Ren-Jie Zhang for interesting
conversations and comments on our manuscript. We also thank
Rong-Gen Cai, Jian-Tao Li for discussions. This project was in
part supported by NNSFC under Grant Nos. 10175070, 10047004 and 
19925523 as
well as also by the Ministry of Science and Technology of China
under grant No. NKBRSF G19990754.


\begin{thebibliography}{99}

\bibitem{S} A. Sen, hep-th/0203211; hep-th/0203265.

\bibitem{GMP} G. Gibbons, hep-th/0204008; S. Mukohyama, hep-th/0204084;
A. Sen, hep-th/0204143;
A. Feinstein, hep-th/0204140;
T. Padmanabhan, hep-yh/0204150;
G. Shiu and I. Wasserman, hep-th/0205003;
T. Padmanabhan and T.R. Choudhury, hep-th/0205055;
S. Mukohyama, hep-th/0208094;
A. Sen, hep-th/0209122;
C. Kim, H.B. Kim and Y. Kim, hep-th/0210101;
H. Lee and W.S. l'Yi, hep-th/0210221.

\bibitem{FC} M. Fairbairn and M.H. Tytgat hep-th/0204070;

\bibitem{CGJP} D. Choudhury, D. Ghoshal, D.P. Jatkar and S. Panda,
hep-th/0204204.

\bibitem{KL} L. Kofman and A. Linde, hep-th/0205121.

\bibitem{PC} Y.S. Piao, R.G. Cai, X.M. Zhang and Y.Z. Zhang, 
Phys. Rev. \textbf{D 66}, 121301(R) (2002); hep-ph/0207143.

\bibitem{G} M.R. Garousi, hep-th/0003122; E.A. Bergshoeff, M. de Roo,
T.C. de Wit, E. Eyras and S. Panda, hep-th/0003221; J. Kluson,
hep-th/0004106.

\bibitem{Se} A. Sen, hep-th/0207105.

\bibitem{ST} A. Sen, JHEP \textbf{9910}, (1999) 008; hep-th/9909062;
A.A. Tseytlin, J. Math. Phys. \textbf{42}, (2001) 2854-2871; hep-th/0011033;
G. Arutyunov, S. Frolov, S. Theisen and A.A. Tseytlin, JHEP \textbf{0102}, 
(2001) 002; hep-th/0012080;
S. Corley, D. Lowe and S. Ramgoolam, JHEP \textbf{0107}, (2001) 030; 
hep-th/0106067;
A. Fotopoulos and A.A. Tseytlin, hep-th/02011101.

\bibitem{M} K. Maeda, Phys. Rev. \textbf{D 39}, (1989) 3159.

\bibitem{SS} E.D. Stewart and D.H. Lyth, Phys. Lett.
\textbf{B 302}, (1993) 171; gr-qc/9302019; J. Hwang and H. Noh,
hep-th/0206100.


\bibitem{STW} G. Shiu, S.-H.Henry Tye and I. Wasserman, hep-th/0207119;
J.M. Cline, H. Firouzjahi and P. Martineau, hep-th/0207156;


\bibitem{GHY} G. Gibbons, K, Hashimoto and P. Yi, JHEP \textbf{0209}, 
(2002) 061; 
hep-th/0209034;
T. Okuda and S. Sugimoto, hep-th/0208196;
N.D. Lambert and I. Sachs, hep-th/0208217.

\bibitem{GS} M. Gutperle and A. Strominger, JHEP \textbf{0204}, (2002) 018; 
hep-th/0202210;
H. Liu, G. Moore and N. Seiberg, JHEP \textbf{0210}, (2002) 031; 
hep-th/0206182;
P. Mukhopadhyay and A. Sen, hep-th/0208142;
A. Strominger, hep-th/0209090;
B. Chen, M. Li and F.L Lin, hep-th/0209232;
Q.G. Huang, K. Ke and M. Li, in preparation.

\bibitem{D} K. Ohta and T. Yokono, hep-th/0207004;
S. Mukohyama, hep-th/0208094;
M. Alishahiha and S. Parvizi, JHEP \textbf{0210} (2002) 047,
hep-th/0208187;
G. Dvali and A. Vilenkin, hep-th/0209217;
E.J. Martinec, hep-th/0210231.



\end{thebibliography}
\end{document}